\documentclass[prl,twocolumn,amsmath,nofootinbib,amssymb,superscriptaddress]{revtex4}

\usepackage{graphicx}

\usepackage{epsfig,psfrag,amsmath,amssymb,float}

\begin{document}

\newcommand{\ltwid}{\mathrel{\raise.3ex\hbox{$<$\kern-.75em\lower1ex\hbox{$\sim$}}}} 
\newcommand{\gtwid}{\mathrel{\raise.3ex\hbox{$>$\kern-.75em\lower1ex\hbox{$\sim$}}}} 
\def\K{{\bf{K}}} 
\def\Q{{\bf{Q}}} 
\def\Gbar{\bar{G}} 
\def\tk{\tilde{\bf{k}}} 
\def\k{{\bf{k}}}

\title{Pairing Glue in the Hubbard and t-J Models}

\author{T.A.~Maier} \affiliation{Computer Science and Mathematics Division,\\
Oak Ridge National Laboratory, Oak Ridge, TN 37831-6164} \email{maierta@ornl.gov}

\author{D. Poilblanc}\affiliation{Laboratoire de Physique Th\'eorique,\\
CNRS \& Universit\'e de Toulouse, F-31062 Toulouse, France} \email{Didier.Poilblanc@irsamc.ups-tlse.fr}

\author{D.J.~Scalapino} \affiliation{Department of Physics,\\
University of California, Santa Barbara, CA 93106-9530} \email{djs@vulcan2.physics.ucsb.edu}

\date{\today} 
\begin{abstract}
	
	The question of whether one should speak of a ``pairing glue" in the Hubbard and $t$-$J$ models is basically a question about the dynamics of the pairing interaction. If the dynamics of the pairing interaction arises from virtual states, whose energies correspond to the Mott gap, and give rise to the exchange coupling $J$, the interaction is instantaneous on the relative time scales of interest. In this case, while one might speak of an ``instantaneous glue", this interaction differs from the traditional picture of a retarded pairing interaction. However, if the energies correspond to the spectrum seen in the dynamic spin susceptibility, then the interaction is retarded and one speaks of a spin-fluctuation glue which mediates the d-wave pairing. Here we present results from numerical studies which provide insight into this question. 
\end{abstract}

\maketitle

%\section{Introduction}\label{sec:1}
The question of whether the pairing interaction in the cuprate superconductors should be characterized as arising from a ``pairing glue" has recently been raised\cite{ref:1}. As we will discuss, this is a question about the dynamics of the pairing interaction and it will be answered when we know more about the frequency dependence of the cuprate superconducting gap. From the d-wave ($\cos k_x-\cos k_y$) momentum dependence of the cuprate gap, we know that the pairing interaction is spatially a short range, dominantly near-neighbor attraction. However, in spite of pioneering ARPES\cite{ref:2,ref:2a,ref:2b,ref:2c,ref:2d}, tunneling\cite{ref:3a,ref:3b,ref:3c} and infrared conductivity\cite{ref:4,ref:4a} studies, we do not yet have sufficient information to definitively characterize its dynamics. Thus, while there is a growing consensus that superconductivity in the high $T_c$ cuprates arises from strong short-range Coulomb interactions between electrons rather than the traditional electron-phonon interaction, the precise nature of the pairing interaction remains controversial.

%\cite{ref:1}
This is the case even among those who agree that the essential physics of the cuprates is contained in the Hubbard and $t$-$J$ models. For example, both Anderson's resonating-valence-bond (RVB) theory\cite{ref:7} and the spin-fluctuation exchange theory\cite{ref:5,ref:6,ref:6a} lead to a short-range interaction which forms $d_{x^2-y^2}$ pairs. However, the dynamics of the two interactions differ. In the RVB picture, the superconducting phase is envisioned as arising out of a Mott-liquid of singlet pairs. These pairs are bound by a superexchange interaction $J$ which is proportional to $t^2/U$. Here $t$ is the effective hopping matrix element between adjacent sites and $U$ is an onsite Coulomb interaction. $J$ is determined by the virtual hopping of an electron of a given spin to an adjacent site containing an electron with an opposite spin.\cite{ref:13} Thus the dynamics of $J$ involves virtual excitations above the Mott gap which is set by $U$, and the pairing interaction is essentially instantaneous. In this case, as Anderson has noted,\cite{ref:1} one should not speak of a ``pairing glue" in the same sense that this term is used when referring to a phonon mediated interaction. In the spin-fluctuation exchange picture, the pairing is viewed as arising from the exchange of particle-hole spin fluctuations whose dynamics reflect the frequency spectrum seen in inelastic magnetic neutron scattering. This spectrum covers an energy range which is small compared with $U$ or the bare bandwidth $8t$. In this case, the pairing interaction is retarded and in analogy to the traditional phonon mediated pairing, one says that the spin-fluctuations provide the ``pairing glue." So the question of whether there is a ``pairing glue" offers a way of distinguishing different pairing mechanisms. Here using numerical techniques we examine this question for the $t$-$J$ and Hubbard models.

In the superconducting state the Nambu self-energy $\hat\Sigma(k,\omega)$ can be parametrized as 
\begin{equation}
	\hat\Sigma(k,\omega)=(1-Z(k,\omega))\omega\tau_0+\chi(k,\omega)\tau_3+\phi(k,\omega)\tau_1. \label{eq:2} 
\end{equation}
Here, $\tau_0$, $\tau_1$ and $\tau_3$ are the Pauli spin matrices and $Z(k,\omega)$ and $\chi(k,\omega)$ describe the so-called normal components of the self-energy and the gap function $\phi(k,\omega)$ describes the anomalous part which contains information on the internal structure of the pairs. The complex gap function $\phi(k,\omega)=\phi_1(k,\omega)+i\phi_2(k,\omega)$ satisfies the Cauchy relation 
\begin{equation}
	\phi_1(k,\omega)=\frac{1}{\pi}\int\frac{\phi_2(k,\omega')}{\omega'-\omega}d\omega' \label{eq:3} 
\end{equation}
and for $\omega=0$, one has 
\begin{equation}
	\phi_1(k,0)=\frac{2}{\pi}\int^\infty_0\frac{\phi_2(k,\omega')}{\omega'}d\omega'\,. \label{eq:4} 
\end{equation}
Based upon this, a useful measure of the frequency dependence of the pairing interaction\cite{ref:13a} is 
\begin{equation}
	I(k,\Omega)=\frac{\frac{2}{\pi}\int^\Omega_0\frac{\phi_2(k,\omega')}{\omega'}d\omega'}{\phi_1(k,0)} \label{eq:5} 
\end{equation}

It gives the fraction of the zero frequency gap function which arises from frequencies below $\Omega$. 
\begin{figure}
	[htb] 
	\includegraphics[width=9cm]{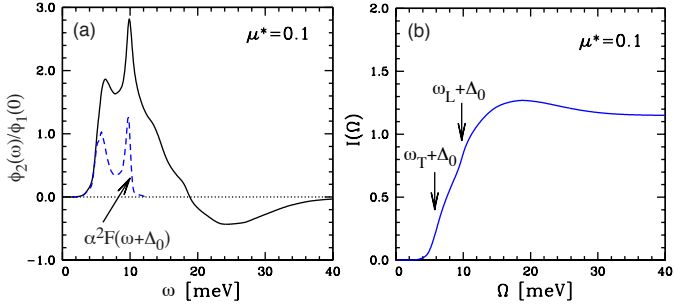} \caption{(a) The imaginary part of the Pb gap function $\phi_2(\omega)$ versus $\omega$ (solid curve). The peaks in $\phi_2(\omega)$ occur at the transverse $\omega_T$ and longitudinal $\omega_L$ peaks of $\alpha^2F(\omega)$ (dashed curve) shifted up by the gap $\Delta_0$. (b) The pairing interaction spectral weight $I(\Omega)$ versus $\Omega$ for Pb. $I(\Omega)$ increases as $\Omega$ passes through $\omega_T+\Delta_0$ and $\omega_L+\Delta_0$ reflecting the transverse and longitudinal phonon contributions to the pairing. At larger values of $\Omega$, $I(\Omega)$ exceeds unity because $\phi_1(0)$ is reduced from the value it would have just due to the phonons by the presence of the non-retarded screened Coulomb pseudo-potential $\mu^*$.} \label{fig:1} 
\end{figure}
In order to obtain some insight into $I(k,\Omega)$, we first consider the case of Pb. Here the $k$ dependence of the gap function is negligible and only the frequency dependence enters. The imaginary part of the gap function $\phi_2(\omega)$, determined from tunneling data\cite{ref:13b} is shown as the solid curve in Fig.~\ref{fig:1}. The dashed curve shows $\alpha^2F(\omega)$. Using this result for $\phi_2(\omega)$ along with the value of $\phi_1(\omega=0)$, we have evaluated $I(\Omega)$. As seen in Fig.~\ref{fig:1}(b), $I(\Omega)$ increases as $\Omega$ passes through the characteristic transverse and longitudinal Pb phonon frequencies plus $\Delta_0$. It then exhibits a broad maximum and settles down to a value that exceeds 1. The maximum arises from the change in sign of $\phi_2(\omega)$ which occurs at a frequency 2 to 3 times the characteristic frequencies of the retarded part of the interaction. The reason that the asymptotic value of $I(\Omega)$ exceeds unity is that the non-retarded screened Coulomb pseudopotential leads to a negative, frequency independent, contribution $\phi_{\rm NR}$ to the real part of $\phi_1(\omega)$. In this case the Cauchy relation Eq.~\ref{eq:3} becomes 
\begin{equation}
	\phi_1(\omega=0)=\frac{2}{\pi}\int^\infty_0\frac{\phi_2(\omega')}{\omega'}d\omega'+\phi_{\rm NR} \label{eq:6} 
\end{equation}
and at high frequencies $I(\Omega)$ exceeds 1 by the non-retarded contribution $-\phi_{\rm NR}/\phi_1(0)$.

%\section{The Hubbard and $t$-$J$ models}\label{sec:3}
The models that we will consider have a square two-dimensional lattice with a near neighbor one electron hopping $t$. The Hubbard model has an onsite Coulomb interaction $U$ and its Hamiltonian is 
\begin{equation}
	H=-t\sum_{\langle ij\rangle s}(c^+_{is}c_{js}+c^+_{js}c_{is})+ U\sum_i n_{i\uparrow}n_{i\downarrow}-\mu\sum_{is}n_{is} \label{eq:7} 
\end{equation}
with $\mu$ a chemical potential which sets the site filling $\langle n\rangle$. Here $c^+_{is}$ creates an electron of spin $s$ on site $i$ and $n_{is}=c^+_{is}c_{is}$ is the site occupation number operator for spin $s$. The $t$-$J$ model is the large $U$ limit of the Hubbard model in which no double site occupancy is allowed and near neighbor spins are coupled by an exchange interaction $J$. 
\begin{equation}
	H=-t\sum_{\langle ij\rangle s}(\tilde c^+_{is}\tilde c_{js}+\tilde c_{js}\tilde c_{is} +J\sum_{\langle ij\rangle}({\bf S}_i\cdot{\bf S}_j-\frac{1}{4}n_in_j) \label{eq:8} 
\end{equation}
Here ${\bf S}_i=\tilde c^+_{is}{\mbox{\boldmath$\sigma$}}_{ss'}\tilde c_{js'}$ and $\tilde c^+_{is}$ is a projected fermion operator defined as $c^+_{is}(1-n_{i-s})$.

Exact diagonalization calculations were carried out for the $t$-$J$ model on a square cluster of $N=32$ sites. This particular cluster exhibits the full local symmetries of the underlying square lattice and has all of the most symmetric $k$ points in reciprocal space. Here we will consider the 0-, 1- and 2-hole sectors. One hole doped on to a 32-site cluster corresponds to a doping $x\simeq0.03$.

%\begin{figure}[htb]
%\includegraphics[width=7cm]{newfig3.eps}
%\caption{The periodic square 32-site $t$-$J$ cluster which was used in the
%Lanczos diagonalization calculations of $G(k,\omega)$ and $F(k,\omega)$.}
%\label{fig:3}
%\end{figure}
The gap function $\phi(k,\omega)$ can be extracted by combining Lanczos results for the one-electron Green's function $G(k,\omega)$ and Gorkov's off-diagonal Green's function\cite{ref:13a,ref:14} 

%\begin{equation}
%	F(k,t)=-i\langle T\tilde c_{-k,-\sigma}(t)\tilde c_{k,\sigma}(0)\rangle. \label{eq:9} 
%\end{equation}
%For a finite cluster, $F(k,\omega)$ is obtained from 
\begin{equation}
	F(k,\omega)=\bar F(k,\omega+i\eta)+\bar F(k,\omega-i\eta) \label{eq:10} 
\end{equation}
with 
\begin{equation}
	\bar F(k,z)=\left\langle\Psi_0(N-2)\right| \tilde c_{-k,-\sigma}\frac{1}{z-H+E_{N-1}} \tilde c_{k\sigma} \left|\Psi_0(N)\right\rangle \label{eq:11} 
\end{equation}
Here the number of electrons in the initial and final groundstates differ by 2 and $E_{N-1}$ is defined as $E_{N-1}=(E_0(N)+E_0(N-2))/2$. For a finite cluster, the diagonal Green's function is defined as 
\begin{eqnarray}
	&&\hspace*{-0.2cm}G(k,\omega)=\nonumber\\
	&&\hspace*{0.2cm}\left\langle\Psi_0(N-2)\right|\tilde c_{k\sigma}\frac{1}{\omega+i\eta-H+E_{N-1}} \tilde c^+_{k\sigma} \left|\Psi_0(N-2)\right\rangle \nonumber \\
	&&\hspace*{0.2cm}+\left\langle\Psi_0(N)\right|\tilde c^+_{k\sigma}\frac{1}{\omega-i\eta+H-E_{N-1}} \tilde c_{k\sigma}\left|\Psi_0(N)\right\rangle \label{eq:12} 
\end{eqnarray}
With this definition, both $G(k,\omega)$ and $F(k,\omega)$ have the same set of energy poles. Using a continued fraction Lanczos based method both $G(k,\omega)$ and $F(k,\omega)$ have been calculated and the gap function $\phi(k,\omega)$ determined from 
\begin{equation}
	\phi(k,\omega)=-\frac{F(k,\omega)}{G(k,\omega)G(k,-\omega)+F^2(k,\omega)}\,. \label{eq:13} 
\end{equation}
\begin{figure}
	[htb] 
	\includegraphics[width=8cm]{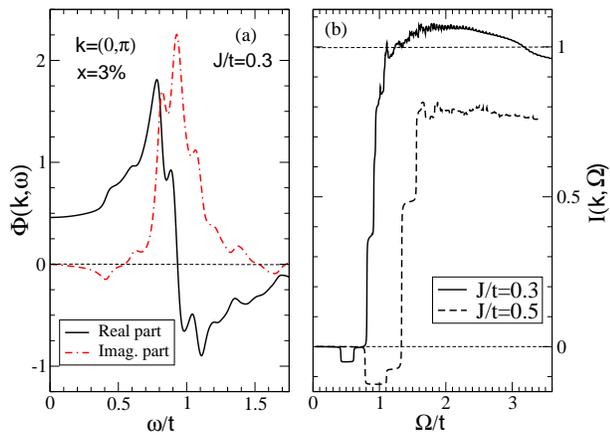} \caption{(a) The real (solid) and imaginary (dashed) parts of $\phi_2(k,\omega)$ versus $\omega/t$ obtained for a 32-site cluster. Here $k=(0,\pi)$, $J/t=0.3$ and the doping $x\simeq3$\%. (b) $I(k,\Omega)$ versus $\Omega/t$ for $J/t=0.3$ (solid) and $J/t=0.5$ (dashed) for $k=(0,\pi)$ and $x\simeq3$\%.} \label{fig:4} 
\end{figure}
Results for $\phi(k,\omega)$ and $I(k,\Omega)$ for $J/t=0.3$ and $x\sim3$\% ($N=32$) with $k=(0,\pi)$ are plotted in Fig.~\ref{fig:4}. We believe that finite size effects are responsible for $\phi_2(k,\omega)$ starting out negatively and that the corresponding negative dip in $I(k,\Omega)$ is an artifact. For $J/t=0.3$, the rapid increase in $I(k,\Omega)$ as $\Omega/t$ exceeds $\sim0.75$ reflects the dynamic contributions of the spin-fluctuations and the broad maximum arises from the negative swing in $\phi_2(k,\omega)$ which occurs when $\omega$ exceeds several times their spectral range. This is similar to the behavior seen in Pb when $\omega$ exceeds the spectral range of $\alpha^2F(\omega)$. At higher frequencies in Fig.~\ref{fig:4}b, $I(\Omega)$ is seen to decrease below 1. This high frequency behavior in which $I(\Omega)$ drops below 1 is more clearly seen for $J/t=0.5$, as shown by the dashed curve in Fig.~\ref{fig:4}b. The fact that at high frequency $I(k=(0,\pi),\Omega)$ lays below 1 means that there is a non-retarded (instantaneous) contribution to the $d$-wave pairing interaction. In contrast to the case of the traditional low temperature superconductors, here the non-retarded contribution increases the pairing, corresponding to a positive value of $\phi_{\rm NR}(k)/\phi(k,0)$. Furthermore its relative contribution increases as $J/t$ increases.

For the Hubbard model, one can explore the full dynamic range including the upper Hubbard band so that the Cauchy relation does not have an additional constant term $\phi_{\rm NR}$. To calculate $\phi(k,\omega)$ for the Hubbard model we have used a dynamic cluster approximation\cite{hettler:dca,ref:8} (DCA). The general idea of the DCA is to approximate the effects of correlations in the bulk lattice with those on a finite size cluster with $N_c$ sites and periodic boundary conditions. The DCA maps the bulk ($L\times L$ with $L\to\infty$) lattice problem onto an effective periodic cluster embedded in a self-consistent dynamic mean-field that is designed to represent the remaining degrees of freedom. The hybridization of the cluster to the host accounts for fluctuations arising from coupling between the cluster and the rest of the system. Here we have used a non-crossing approximation\cite{ref:8,maier:dcanca} (NCA) to determine $\phi(k_A,\omega)$ for a 4-site $2\times2$ cluster at a wave vector $k_A=(0,\pi)$. This cluster allows for a gap with $d$-wave symmetry and is such that within the non-crossing approximation dynamic results can be obtained on the real frequency axis. Similar calculations were performed for the t-J model in Ref.~\cite{haule07}.

In mean-field theories such as the DCA, the mean-field generates a constant real term $\phi_{\rm MF}(k_A)$. In the infinite cluster size limit, the DCA recovers the exact result and the mean-field contribution $\phi_{\rm MF}(k_A)$ vanishes. For a finite cluster size, we therefore view this contribution as an artifact and subtract it off of $\phi_1(k_A,\omega)$ before performing the analysis based on the Cauchy relation. $\phi_{\rm MF}(k_A)$ was determined from $\lim_{\omega\to\infty} \phi(k_A,\omega)$. The expression for $I(\Omega)$ becomes 
\begin{equation}
	I(k_A,\Omega)=\frac{ \frac{2}{\pi}\int^\Omega_0\frac{\phi_2(k_A,\omega')}{\omega'}d\omega'}{ \frac{2}{\pi}\int^\infty_0\frac{\phi_2(k_A,\omega')}{\omega'}d\omega'}= \frac{ \frac{2}{\pi}\int^\Omega_0\frac{\phi_2(k_A,\omega')}{\omega'}d\omega'}{ \phi_1(k_A,0)-\phi_{\rm MF}(k_A)}\,. \label{eq:14} 
\end{equation}

\begin{figure}
	[htb] 
	\includegraphics[width=8cm]{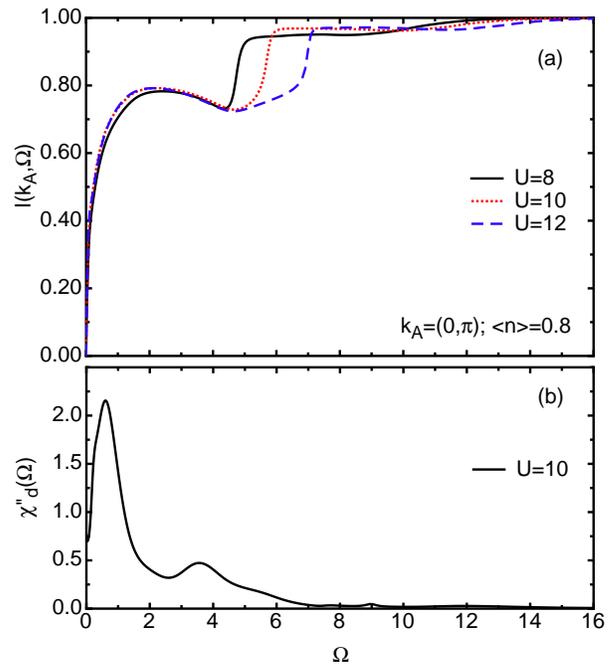} \caption{(a) $I(k_A,\Omega)$ versus $\Omega/t$ for the $2\times2$ DCA-NCA Hubbard calculation with $U/t=10$, $\langle n\rangle=0.8$ and  $T/T_c\simeq 0.95$. Here $k_A=(0,\pi)$. (b) The $d$-wave projected $\chi''_d(\Omega)$ versus $\Omega/t$ for the same parameters.} \label{fig:5} 
\end{figure}

Results showing $I(k_A,\Omega)$ versus $\Omega$ for a filling $\langle n\rangle=0.8$ and $U/t=10$ are plotted in Fig.~\ref{fig:5}a. The $d$-wave projection of the dynamic spin susceptibility 
\begin{equation}
	\chi''_d(\Omega)=\frac{\langle(\cos k'_x-\cos k'_y)\chi''(k-k',\Omega)(\cos k_x-\cos k_y)\rangle} {\langle(\cos k_x-\cos k_y)^2\rangle} \label{eq:15} 
\end{equation}
was calculated for $U/t=10$ and $\chi''_d(\Omega)$ is shown in Fig.~\ref{fig:5}b. These calculations are for a reduced temperature $T/T_c\simeq0.95$. Once $T<T_c$, the $\tau_1$ component of the Nambu self-energy gives $\phi(k_A,\omega)$ and one can calculate $I(k_A,\Omega)$. For $T/T_c=0.95$ the shift due to the magnitude of the gap at the antinode $\Delta(k_A)$ is negligible. 

\begin{figure}
	[htb] 
	\includegraphics[width=8cm]{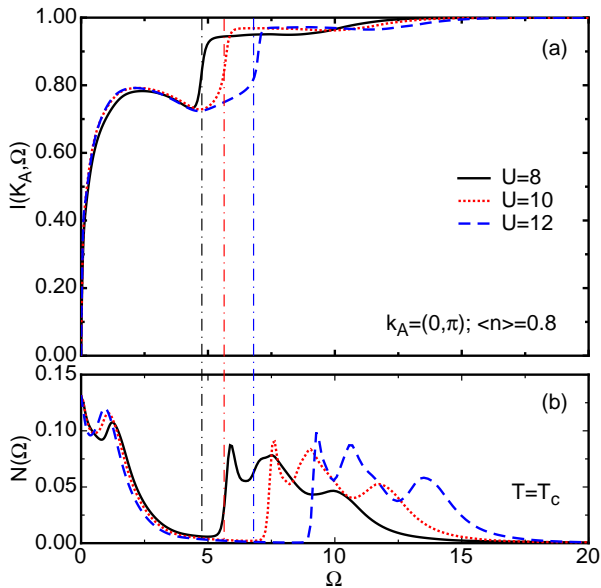} \caption{(a) $I(k_A,\Omega)$ for the $2\times2$ DCA-NCA Hubbard calculations with different values of $U/t$ at a filling $\langle n\rangle=0.8$ and $T/T_c\simeq 0.95$. (b) The density of states $N(\Omega)$ at $T=T_c$ for $\langle n\rangle=0.8$ and different values of $U/t$. Here one sees the increase in the Mott gap that separates the upper and lower Hubbard bands.} \label{fig:4} 
\end{figure}

At low frequencies $I(k_A,\Omega)$ is seen to increase over the spectral range associated with the spin-fluctuation response seen in $\chi''_d(\omega)$. As $\Omega$ exceeds this range, $I(k_A,\Omega)$ passes through a weak maximum and dips down slightly, similarly to the $t$-$J$ results shown in Fig.~\ref{fig:4}b. Then however, on a higher energy scale $I(k_A,\Omega)$ goes to 1. Similar results for $I(k_A,\Omega)$ for different values of $U/t$ are shown in Fig.~\ref{fig:4}a. Here one sees that as $U/t$ increases, the region over which $I(k_A,\Omega)$ remains below 1 extends to higher energies. For the $t$-$J$ model this energy was pushed to infinity, but for the Hubbard model the high frequency contribution that takes $I(k_A,\Omega)$ to 1 is associated with the upper Hubbard band as seen from the single-particle density of states shown in Fig.~\ref{fig:4}b.

%\begin{figure}[htb]
%\begin{center}
%\includegraphics[width=7cm]{fig7.eps}
%\includegraphics[width=9cm]{d-wave-projec-im_chi.eps}
%\end{center}
%\caption{(a) $I(k_A,\omega)$ Hubbard results for different values of $U/t$ and
%doping obtained from a VCA calculation using an 8-site cluster. (b) Results for
%the imaginary part of the $d$-wave projected magnetic spin susceptibility.}
%\label{fig:7}
%\end{figure}
%
%\section{Conclusions}\label{sec:4}
These numerical results show that the $d$-wave pairing interaction in the $t$-$J$ and Hubbard models contain both retarded and non-retarded contributions.\cite{ref:15} The retarded contribution occurs on an energy scale which is small compared to the bare bandwidth $8t$ and the onsite Coulomb interaction $U$. For the Hubbard model, the ``non-retarded" contribution occurs on an energy scale set by the Mott gap and is related to excited states involving the upper Hubbard band. For the $t$-$J$ model, this energy scale is pushed to infinity and the exchange contribution is instantaneous.

A simple phenomenological form for the $d$-wave pairing interaction, consistent with these observations is 
\begin{equation}
	\frac{3}{2}\bar U^2\chi(k-k',\omega,\omega')-\bar J(\cos k_x-\cos k_y)(\cos k'_x-\cos k'_y)\,. \label{eq:16} 
\end{equation}
Here $\chi(q,\omega)$ is the dynamic spin susceptibility and $\bar U$ and $\bar J$ are effective coupling constants. The retarded contribution to the pairing comes from the first term, and the non-retarded contribution from the second, exchange, term.
Unlike the traditional low $T_c$ case where the non-retarded screened Coulomb interaction suppresses the gap, here the non-retarded exchange term enhances the $d$-wave gap.

The question regarding whether there is a ``pairing glue" is then a question of whether the dominant contribution to $\phi_1(k_A,\omega=0)$ comes from the integral of $\phi_2(k_A,\omega)/\omega$.\cite{ref:comment} From the results presented here we conclude that both the $t$-$J$ and Hubbard models have spin-fluctuation ``pairing glue". However, they also exhibit a smaller, non-retarded contribution. For the cuprate materials, the relative weight of the retarded and non-retarded contributions to the pairing interaction remains an open question. Thus the continuing experimental search for a pairing glue in the cuprates is important and will play an essential role in determining the origin of the high $T_c$ pairing interaction.

\section*{Acknowledgments} We would like to thank E.~Nicol for her assistance with Figure~\ref{fig:1} and R.~Melko for his help with the McMillan-Rowell data of Reference~18. DJS would like to thank W. Hanke and M. Aichhorn for useful discussions and acknowledge the Aspen Center for Physics where the question of pairing glue was actively discussed. This research was enabled by computational resources of the Center for Computational Sciences at Oak Ridge National Laboratory and the French supercomputer center IDRIS at Orsay (France). TAM and DJS acknowledge the Center for Nanophase Materials Sciences, which is sponsored at Oak Ridge National Laboratory by the Division of Scientific User Facilities, U.S. Department of Energy.

\end{document}